\documentclass[a4paper,10pt]{article}

\usepackage{epsfig}

\begin{document}

\begin{center}
\noindent{\Large \bf Testing entropy production hypotheses in non-linear steady states} \\

\vspace{15pt}

Stijn Bruers\footnote{email: stijn.bruers@fys.kuleuven.be, tel: 0032(0)16327503}\\
Instituut voor Theoretische Fysica, Celestijnenlaan 200D,\\
Katholieke Universiteit Leuven, B-3001 Leuven, Belgium\\
\end{center}

\begin{abstract}
In the last few decades, some hypotheses for entropy production (EP) principles have been forwarded
as possible candidates for organizational principles in non-linear non-
equilibrium
systems. Two important hypotheses will be studied: the maximum entropy production (MaxEP) principle that claims that the selected steady state has the highest EP, and the gradient response principle that claims that the EP of the selected steady state (maximally) increases when the external thermodynamic driving force increases. We will formulate these hypotheses more rigorously and present a simple chemical reaction model to test these hypotheses.
With the help of this model, we will clearly demonstrate that there are
different MaxEP hypotheses being discussed in the literature and we will look at
some parts in the literature where these differences are not always clarified.
Furthermore, our chemical model will be
a general counter example to all of these MaxEP and gradient response hypotheses.
\end{abstract}

\footnotesize
\noindent {\bf pacs numbers:} 05.70.Ln, 65.40.Gr, 82.60.-s \\
{\bf KEY WORDS:} maximum entropy production, nonequilibrium thermodynamics,
chemical reactions, variational principles 
\normalsize


\section{Introduction}

In the linear response regime near thermodynamic equilibrium, it is well known
that one can derive constitutive equations of motion by using a variational
principle of least dissipation (Onsager, \cite{Onsager, OnsagerMachlup}), and
one can find the unique steady state by minimizing the entropy production
(MinEP) under some physical constraints (as first discussed by Prigogine
\cite{Prigogine, KondepudiPrigogine}). By using other constraints, one can find
respectively the constitutive equations of motion and the steady state by
maximizing the entropy production (MaxEP, see resp. \cite{Ziegler} and \cite{Ziman,
ZupanovicJuretic2004}).

The situation far from thermodynamic equilibrium, with nonlinear 
dynamics and non-linear response, is much more difficult. Far from equilibrium
not only involves thermodynamic 
constraints, but the description is also highly dependent on the 
kinetics
(the balance or constitutive equations). Entropy 
production (EP) is a fundamental notion in irreversible 
thermodynamics,
because it combines entropy (thermodynamics) with time (kinetics), 
it is tempting to look for EP principles. Besides successes of the
near-equilibrium (linear response)
MinEP and MaxEP, there is also some renewed interest in a non-linear MaxEP, for
systems far from equilibrium\footnote{With systems far from equilibrium, we mean
systems in local equilibrium, but not in the linear response regime. Although
(non-) linear response is not equivalent with (non-) linear dynamics, the
referred studies as well as the model in this article are non-linear in both
meanings.} (see 
e.g.
\cite{KleidonLorenz}), from complex chemical 
reaction
systems \cite{JureticZupanovic} to fluid systems \cite{ShimokawaOzawa} 
or even ecological \cite{Kleidon2004} and climate systems
\cite{Paltridge1975}. Recent reviews include \cite{MartyuchevSeleznev2006} and
\cite{OzawaOhmura}.

Schneider et al. \cite{SchneiderKay} also formulated another EP hypothesis, which we will call the gradient response principle. The above MinEP and MaxEP principles are formulated for a fixed external thermodynamic driving force (external applied gradient), whereas the gradient response studies the behavior of the EP in the selected steady state, when the external gradient increases. The intuition behind this principle is that when systems are pushed further away from equilibrium, they will try harder to get to equilibrium. As the external applied gradient is a measure for the distance from equilibrium and the EP is a measure for how strong the system tends to equilibrium, the gradient response hypothesis roughly states that the steady state EP (maximally) increases when the external applied gradient increases. 

In this article, a chemical reaction model is presented, which is 
inspired by a resource-consumer-predator system in ecology. The system's EP
properties in the steady states are studied, and it is used to test the EP hypotheses.
With this model we can easily demonstrate that there are actually different
MaxEP principles in use in the literature. Apart from the distinction between linear and non-linear principles, one can make a distinction between steady state principles and transient principles. The latter principles are used e.g. to derive constitutive equations of motion which are valid in both the transient and the steady states. An example is Onsager's least dissipation \cite{Onsager, OnsagerMachlup}. Also Prigogine's MinEP principle \cite{Prigogine} compares the steady state EP with the EP in the neighboring transient states. The latter principle (not to be confused with Onsager's) can also be called a Lyapunov principle because in this principle the EP is mathematically a Lyapunov functional \cite{KondepudiPrigogine}. 

We will only examine the steady state non-Lyapunov principles, which we have named the partial steady state MaxEP, the
non-variational MaxEP and the maximum gradient response for reasons that will become clear later.
Some parts in the literature where these differences between MaxEP
principles are not clarified, will be mentioned in the final discussion section.
These non-linear MaxEP principles are also very different from the linear MaxEP
principles \cite{Ziegler, Ziman, ZupanovicJuretic2004} (which are basically
correct in the linear response regime).

Apart from pointing at these differences between MaxEP principle, a second important
result is that this one chemical reaction model might serve as a general counter
example to all of these most used MaxEP hypotheses (although some hypotheses have much simpler counter examples). The author is not aware of real systems obeing the same dynamics, and therefore references to experimental studies will not be made. From a theoretical point of view, the model is consistent and obeys all known physical laws. It is not more complex than e.g. the Belousov-Zhabotinsky system \cite{KondepudiPrigogine}. If nature in reality excludes all counter example systems, this would be a new physical law.

\section{The chemical reaction model}

\subsection{General description}

Let us study the entropy production in a specific chemical reaction 
system. The system consists of five chemical substances $A,\, R,\, 
C,\, P$
and $W$. The reaction set has six reactions
\begin{eqnarray}
A &\rightleftharpoons& R,\\
2R+C &\rightleftharpoons& 2C+W,\\
2R+P&\rightleftharpoons& 2P+W,\\
2C+P&\rightleftharpoons& 2P+W,\\
C&\rightleftharpoons&W,\\
P&\rightleftharpoons&W.
\end{eqnarray}
The concentrations will be denoted with the same letters. The 
concentrations $A$ and $W$ are kept fixed. This reaction scheme is well known
and much studied in ecology as a description for a resource-consumer-predator
ecosystem \cite{MyliusKlumpers}, whereby $A$ and $R$ represent the resource, $C$
is the primary consumer, $P$ is the (omnivore) predator and $W$ is the dead
organic waste. 

Each reaction has a rate $F_i$ ($i=1,\,..6$), and they determine the dynamical equations. We will take them as simple as possible, but still physically realistic: 
\begin{eqnarray}
\frac{d R}{d t}&=&
f_{AR}A-b_{AR}R-2f_{RC}R C -2f_{RP}RP, \label{dynR}\\
\frac{d C}{d t}&=&
f_{RC}R C-2f_{CP}CP-f_
{C}C, \label{dynC}\\
\frac{d P}{d t}&=& f_{RP}RP+f_{CP}CP-f_{P}P.\label{dynP}
\end{eqnarray}
We have neglected the (backward rate) terms in $W$ because this concentration is assumed to be very small.

The thermodynamic forces $X_i$ for each reaction are given by the affinities, the sum of the chemical
potentials, weighted by their stoichiometric coefficients \cite{KondepudiPrigogine}. For ideal gases or ideal and perfect solutions, the chemical potentials are (up to constants) given by the logarithm of the concentrations. We will 
only need the following overall affinities:
\begin{eqnarray}
X_{AW}&=&\mu_A-\mu_W = \ln\frac{K_{AW}A}{W},\\
X_{AR}&=&\mu_A-\mu_R = \ln\frac{K_{AR}A}{R},\\
X_{RW}&=&\mu_R-\mu_W = \ln\frac{K_{RW}R}{W},
\end{eqnarray}
with $X_{AR}+X_{RW}=X_{AW}$ the total, external driving force (applied gradient), which is fixed. The latter equation leads to the relation $K_{AW}=K_{AR}K_{RW}$ between the equilibrium 
constants. Note that the $W$ can not be neglected in the logarithms and that for simplicity we have neglected the absolute temperature factor in front of the logarithm.

The total EP can be written as the sum of the six terms $\sigma_{tot}
=\sum_i F_iX_i$. After some calculations and some thermodynamic 
consistency
equations (basically Hess's law, see e.g. \cite{KondepudiPrigogine}) 
one can write down a very simple expression for the total steady 
state EP (the steady states are denoted with upperindex $\gamma$):
\begin{eqnarray}
\sigma_{tot}^\gamma&=&(f_{AR}A-b_{AR}R^\gamma)\ln\frac{K_{AW}A}{W}.
\end{eqnarray}
This can be understood by observing that there is no net accumulation of $R$, $C$ or $P$, and therefore the overall
reaction rate from $A$ to $W$ is $f_{AR}A-b_{AR}R^\gamma$. This should be multiplied with the overall force $X_{AW}$.

Next, we have to solve the dynamics, find the steady states 
(especially $R^\gamma$), and determine the asymptotic stability. The 
dynamics looks like a resource-consumer-omnivore 
ecosystem model, whose steady states and stability were calculated in \cite{MyliusKlumpers}. 

\begin{figure}[!ht]
\centering
\includegraphics[scale=0.6]{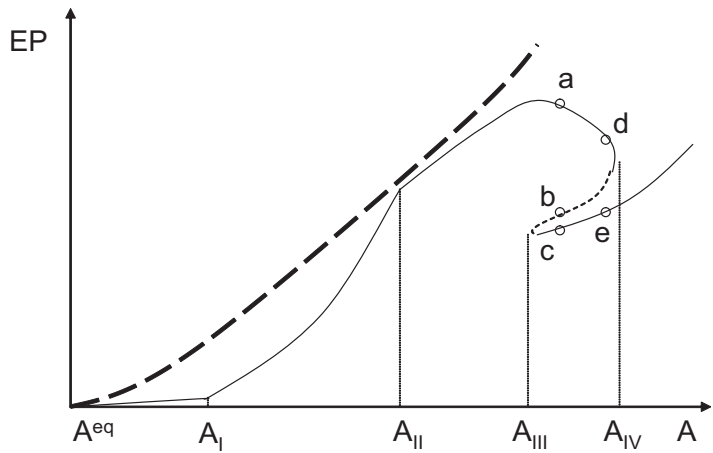}
\caption{A qualitative picture of the EP in the steady states as a 
function of $A$. The thick dashed line is the EP obtained from the
partial steady state MaxEP principle. The thin solid lines are the EP in the 
asymptotically stable steady states. The thin dotted line consists 
of
the unstable states.} \label{EPcounterexamplefig}
\end{figure}

We will not present the expressions for the steady states here, but we will
immediately present a qualitative picture (by taking the parameter 
values as e.g. $f_{AR}=b_{AR}$ $=f_{CP}=f_{C}=1$, $f_{P}=f_{RC}$
$=f_{RP}=2$, and doing some rescaling to make things more visible) of the total steady state EP in Fig. \ref{EPcounterexamplefig}. As the overall constant driving force $X_{AW}$ is an increasing function of the parameter $A$ (for constant and very small $W$), it is sufficient to express the EP as a function of $A$. Only the realistic 
steady states are shown; there are other unphysical 
states with (very small)
negative concentrations. Calculating the stability is simply done by 
looking at the eigenvalues of the dynamical system linearized around 
the
steady state. There are two saddle-node (fold) bifurcations at $A_{III}$ and $A_{IV}$. The states indicated by 'c' and 'd' are on the stable branches, but when they get close to the saddle-node points, their linear stability decreases. Therefore, by taking them sufficiently close to the saddle-nodes, they have a lower stability than the states 'a' and 'e' which are further away from the saddle-nodes. This remark will become important in our later discussions.

With this set-up, we can look at MaxEP. We will see that there are 
different MaxEP principles. These differences will become clear by 
using our
chemical reaction example.

\subsection{Partial steady state MaxEP} \label{Variational MaxEP}

The work by Paltridge \cite{Paltridge1975, Paltridge1978, Paltridge1979} to
understand the convective heat flows in the earth 
atmosphere was a
starting point to study maximum entropy production in highly non-
linear systems. The basic idea behind the Paltridge model is rather 
simple. The atmosphere is divided in two compartments, the equator and the poles, and only the energy balance in the atmosphere is considered.
Solar energy is irradiated at the equator. There is an atmospheric 
(and oceanic)  heat flux from the equator to the poles, where the 
energy is
reradiated back into space. All the processes can be split into 'simple' or linear
and 'complex' or non-linear ones. In the Paltridge model, the simple processes are basically the 
radiation
processes, the complex processes are  
the heat transport processes by fluid convection from equator to pole. The
non-linear highly complex atmosphere subsystem is regarded as a black box, without
knowing the exact internal dynamics. MaxEP claims that the heat transport 
coefficient of the atmosphere, the heat flow from equator to pole, and 
the
driving force (the temperature gradient) on earth will settle 
themselves in a state of maximum atmospheric EP. 
This partial steady MaxEP approach has been made more precise and extended to study
atmospheres of
other planets
\cite{Grassl, KleidonLorenz, LorenzLunine, OhmuraOzuma, OzawaOhmura,
WyantMongroo}. The derived values for the steady state heat 
transport
coefficient are consistent with a number of experiments. 

Remarkably, this MaxEP principle was later on
also applied to other physical systems, like electric arcs \cite{Christen}, 
photosynthetic \cite{JureticZupanovic} or ATP synthase \cite{DewarJuretic} chemical reactions. In these
chemical reaction
system, there were again 'simple' and 'complex' reactions. The 'complex'
chemical reaction rate parameters (e.g. the parameters regulating the transition P.ADP$\rightleftharpoons$ATP in \cite{DewarJuretic}) were derived by postulating MaxEP, 
and
these obtained values were also compatible with experimental data.

Stated generally, the partial steady state MaxEP
principle states that the EP in the steady state of only the complex non-linear processes, i.e. a \emph{partial} EP (not the total EP of all processes), is 
maximized with respect to a continuous 'effective' parameter (e.g. the effective atmospheric heat transport or the reaction rate parameter) or 'effective' flux related with those complex processes. 

To make the procedure more clear, let us apply this partial steady state MaxEP principle to our model. As can be seen,
$F_{AR}=f_
{AR}A-b_{AR}R$ is a 'simple' linear expression. The non-linearities occur in 
the flow
from $R$ to $W$. First we will describe the latter flow as a 'black box',
without specifying the internal dynamics. Afterwards, we will compare the
obtained MaxEP result with the internal dynamics (\ref{dynR}-\ref{dynP}). 

Suppose the system selects a steady state, which we will write with an
upperindex $*$. The MaxEP principle claims that this selected state is the one
which has a maximum EP for the non-linear processes. These processes can be grouped in an 'effective' flux $F_{RW}^*=f_{RW}R^*$ from $R$ to $W$ (again the term containing $W$ is neglected in the flux expression), with $f_{RW}$ the 'effective' parameter. The non-linear part of the
EP is given by
\begin{eqnarray}
\sigma^*_{RW}&=&F^*_{RW}X^*_{RW}\\
&=&F_{RW}^*\ln\frac{K_{RW}R^*}{W}\\
&=&F_{RW}^*\ln\frac{K_{RW}(K_{AR}A-F^*_{RW}/b_{AR})}{W},
\end{eqnarray}
with $0\leq F_{RW}^*\leq K_{AR}b_{AR}A$. We have used the steady 
state constraint $F_{RW}^*=f_{AR}A-b_{AR}R^*$ and $K_{AR}=f_{AR}/b_{AR}$.

As one can see, in the above expression there is a trade-off between 
$F^*_{RW}$ and $X^*_{RW}$. As the former increases, the latter 
decreases,
and vice versa. Therefore, there is an optimal value for $F_{RW}^*$. 
Taking the maximum of this EP with respect to $F_{RW}^*$, one finds
\begin{eqnarray}
F_{RW,MaxEP}^*=\frac{\left( \mathbf{L_W}(K_{AW}A e/W)-1\right)K_{AR}b_
{AR}A}{\mathbf{L_W}(K_{AW}A e/W)}
\end{eqnarray}
with $\mathbf{L_W}$ the LambertW function and $e$ Euler's number. The 
above expression gives $F_{RW}^*$ as a function of $A$ (at constant and very small $W$).

With this expression we can calculate $R_{MaxEP}^*$ and the total EP, which is
indicated in Fig.
\ref{EPcounterexamplefig}. Let us now compare this result with the EP obtained
from the specified internal dynamics (\ref{dynR}-\ref{dynP}). It is clear that
this does not correspond 
with the EP's in the stable steady states by directly solving the concrete
dynamics. We can conclude that the
partial steady state MaxEP principle does not hold in our example.

There is a difference between the atmospheric system and our chemical system.
The atmosphere is highly non-linear and has a lot of possible 
processes and degrees of freedom. One can argue that even though our chemical
system is non-linear, it 
is not
'non-linear enough', or it does not have enough degrees of freedom 
(we have basically only three macroscopic variables: $R,\, C$ and $P$). Keeping
the successes of equilibrium thermodynamics in mind, some intuition might
indicate that a similar kind of 'law of large numbers' can be applied in
non-equilibrium complex systems, resulting in MaxEP. However, it is not clear
why adding macroscopic variables or other non-linear terms should result in the
correct flow 
$F_{RW,MaxEP}^*$. The steady state solutions remain to sensitive on
the parameter values. To explain the experimental successes of MaxEP in e.g.
atmospheric models, one need to know what kind of properties are required for
the internal dynamics in order to obtain the correct MaxEP state.

\subsection{Non-variational MaxEP}

It is very often claimed \cite{MartyuchevSeleznev2006, Sawada1984, 
Shapharonov, Swenson} that an isolated non-equilibrium system relaxes 
to
equilibrium 'as fast as possible', that it 'follows the most efficient 
route' to increase its entropy \cite{Woo} or that it 'selects the path'
with highest EP. The idea behind this is that a non-equilibrium system 
is in a small region in phase space with low entropy, and has the 
highest
probability to evolve in the next time step to the largest region, 
with highest entropy. These statements are still quite vague, and it 
is not
clear what kind of constraints are involved.

One can try to make the above formulations a bit more precise. First 
we have to make the notion of 'paths' more precise. Let us restrict the
'paths' to (pseudo) steady states: Suppose there are different 
paths, i.e. steady states, then the steady state with highest 
EP,
i.e. the one moving to equilibrium as fast as possible, will 
be 'selected'. Secondly we have to make the notion 'select' more 
precise. The real
steady state that is selected can be the one that is e.g. the most 
stable. There are many notions of stability. Let us take asymptotic
(linear) stability. The reformulation reads: 'The steady state 
with highest EP is the most asymptotically stable.' Mathematically, 
this
means that when the dynamics are highly non-linear, then they might 
lead to different steady states $M^\gamma$ ($M=A,\,R,...,W$) and
$X_i^\gamma (M^\gamma),\, F_i^\gamma(M^\gamma)$. Denote with upper 
index $\gamma=*$ the most asymptotically stable state. Then the claim is
\begin{eqnarray}
\sigma^*(X^*,F^*)\geq \sigma^\gamma(X^\gamma,F^\gamma),\qquad \forall 
\gamma.
\end{eqnarray}

An important remark is that this principle is not a variational 
principle, because there is no action and no variation with 
respect to
continuous variables (such as fluxes) or 'effective' parameters. It is rather a selection 
principle of a discrete number of steady states. In this sense, it 
is from a
very different nature than the non-linear partial steady state MaxEP.

This non-variational MaxEP principle is also related to the notion of 
dissipative systems with dissipative structures. If one drives the 
system
out of equilibrium, at certain critical levels of the driving force, 
bifurcations to other stable states are possible. Then a patterned or
ordered structure might arise. A famous example is the Rayleigh-
B\'enard system \cite{Rayleigh}. This consists of a viscous fluid 
subject to a
gravitating field and a temperature gradient: The bottom layer is 
heated whereas the upper is cooled. At a critical level of the 
temperature
gradient, the heat-conducting state is transformed to a heat-
convecting state, with convection cells in a regular pattern, called 
the
dissipative structure. The claim is that this ordered dissipative 
structure (if it exists) always has a higher EP than the so called
'thermodynamic branch' state without the structure, i.e. the state, 
like the conduction state, which do not show a pattern.

There is some verification of this principle from a number of studies. 
The most important field to study this principle is fluid dynamics.
Shimokawa et al. \cite{OzawaShimokawa, ShimokawaOzawa}, based on work 
by e.g. Malkus \cite{Malkus}, studied turbulent and (oceanic) fluid
systems, and they discovered that the MaxEP state is most stable 
against perturbations. Also Schneider et al. \cite{SchneiderKay} 
describe the
increase in EP when the B\'enard fluid system system moves to the 
stable convection state. Renn\'o \cite{Renno} suggested that the most 
stable
state in a radiative-convective atmosphere model with two stable 
states has the highest EP. Also in the Brusselator chemical reaction 
system,
non-variational MaxEP was observed with numerical simulations (see 
Sawada \cite{Sawada1984}, although Sawada termed it perhaps 
confusingly a
variational principle, although there was not a clear notion of an action
presented.)

Other studies showed possible counterexamples to the non-variational 
MaxEP principle. When the external driving force parameter increases,
bifurcations towards new patterns and dissipative structures might 
occur. Most of the above mentioned studies were restricted to the 
dissipative
structures after the first bifurcation (e.g. the transition from the 
conduction to the convection state). However, when the system is pushed
further out of equilibrium, new bifurcations might arise, resulting 
into new stable states and patterns. The old states become unstable. And as is
shown by numerical simulations
\cite{CastilloHoover, CleverBusse, Nicolis}, the total heat transport 
and EP of these new states might be \emph{lower} as compared with the unstable
states. (Nicolis
\cite{Nicolis} also gave a counterexample of the non-variational MaxEP 
principle for chemical reactions.)

These numerical counterexamples were criticized by Martyuchev et. al. 
\cite{MartyuchevSeleznev2006} by claiming that there are computational
difficulties involved, that the criteria of stability and coexistence 
in numerical simulations are subjective, and that taking time-averages 
or
spatial integrations are dubious. Our chemical reaction model can 
serve as a simple counterexample for the non-variational MaxEP 
principle,
without the need for computer simulations because it is analytically 
solvable. Hence the above criticism does not apply to our case.

Let us go to our model. As can be seen in Fig. \ref{EPcounterexamplefig}, there
are two critical levels $A_I$ and $A_{II}$, with sharp changes in the EP. The
behavior near $A_I$ is analogous to the bifurcation behavior in the
Rayleigh-B\'enard convection system, switching from the conduction state (for
$A<A_I$) to the convection state (when $A>A_I$). The interesting property of our
chemical reaction system is that we can easily look what happens after a second
bifurcation at $A_{II}$, without the need for numerical simulations. As discussed above, the state corresponding 
with 'd' in Fig. \ref{EPcounterexamplefig} is \emph{less} stable but 
has a
\emph{higher} EP than the state in 'e': $\sigma^d > \sigma^e$. This counters the
non-variational MaxEP hypothesis.

This counterexample was perhaps already hinted at by Sawada \cite
{Sawada1984}, who claimed that the non-variational MaxEP could be 
wrong in the
presence of hysteresis. Nevertheless, there are still the numerical 
Rayleigh-B\'enard system simulations without non-variational MaxEP and
without hysteresis \cite{CastilloHoover}.




\subsection{Maximum gradient response}

Schneider and Kay \cite{SchneiderKay} studied the degradation of an externally 
applied gradient. This gradient is the external driving force, and its
degradation means an EP. Schneider and Kay looked at the change in EP 
when the external driving force is increased. They formulated what they
have called a 'restated second law of thermodynamics'.\\
"The thermodynamic principle which governs the behavior of systems is 
that, as they are moved away from equilibrium, they will utilize all
avenues available to counter the applied gradients. As the applied 
gradients increase, so does the system's ability to oppose further
movement from equilibrium."\\
We have called this principle the maximum gradient response. It needs 
some further specification, because it is still quite vague.

We can give at least three different interpretations. These interpretations are
formulated as: In the most asymptotically stable state\\
-the EP is positive.\\
-the EP is increasing as the gradient increases.\\
-the EP is increasing and it is a steeper function compared with the EP of the
less 
stable steady states.

The first of the above statements is nothing but the second law. The latter two
statements are a weaker and a stronger extension. Our model can serve as a
counterexample of 
these
latter two statements. 

As mentioned, the state 'a' in Fig. \ref{EPcounterexamplefig} is more asymptotically stable than 'b' (which is 
unstable) or 'c'.
However, $\frac{d \sigma^a}{d A}\leq 0$ and $\frac{d \sigma^b}{d A}
>\frac{d \sigma^a}{d A}$. So the most stable steady state is not
always increasing, nor is it the steepest.

Related with this gradient response principle, Woo \cite{Woo} 
discussed the behavior under pseudo-stationarity conditions: When the 
external
reservoirs (corresponding with $A$ and $W$) are very large but finite, 
the concentrations $A$ and $W$ are not fixed but they are very slowly
relaxing towards equilibrium. It was claimed that $\frac{d \sigma^*}{d 
\tau}\leq 0$, with $\tau$ the time corresponding with the time scale of
this relaxation process. This relaxation is basically a slow movement 
towards the equilibrium value $A^{eq}=W/K_{AW}$ far left in the Fig.
\ref{EPcounterexamplefig}. However, a movement from 'd' to 'a' is 
possible, leading to an \emph{increase} in EP, instead of a decrease. 
Even
sudden decreasing or increasing jumps are possible when $A$ is varied.

\section{Further discussion}

As we have seen, our chemical reaction model clearly shows a counter argument
for MaxEP. Looking at Fig. \ref{EPcounterexamplefig}, one can argue that 'anything
goes' for the EP, except that the specific EP for every independant reaction is
positive.

But what we can also remark, is that there is not one MaxEP principle, but there
are different principles, having very different descriptions. The distinctions
between these MaxEP principles, especially between the partial steady state MaxEP and the
non-variational MaxEP, are not always clear in the literature. We will briefly
discuss some of these shortcomings in the literature.

In \cite{Woo}, the non-variational MaxEP is misleadingly related with a
variational principle, the least dissipation. The latter principle by
\cite{Onsager} is only valid near equilibrium, i.e. in the linear response
regime. The possibility for a non-linear least dissipation principle is still unknown. Nevertheless, the connection in \cite{Woo} between a variational and
a non-variational principle was proposed, without stressing its differences. 

Also in e.g. Dewar \cite{Dewar2005} and Ozawa et al. \cite{OzawaShimokawa}, as
well as in some reviews \cite{MartyuchevSeleznev2006} and \cite{OzawaOhmura},
both the partial steady state and non-variational principles are discussed without
stressing their differences. The discussion of the partial steady state principle is
mostly done by using the atmospheric climate system, whereas the discussion of
the non-variational principle mostly uses the Rayleigh-B\'enard convection
system. However, as we have seen, these principles are very different, because
for example there is no guarantee that the most stable state (with highest total
EP) has the correct value for some transport coefficient (e.g. the heat flow
rate in the atmosphere, or a chemical reaction rate) such that the partial EP
related with this transport is maximal.

One can also see the difference between partial steady state and non-variational MaxEP as
follows. The non-variational MaxEP uses the total steady state EP. This can
be written as $\sigma^*=\bar X F^*$ with $\bar X$ the constant overall force
(e.g. the fixed boundary temperature difference in the Rayleigh-B\'enard system,
or the fixed chemical potential difference in the chemical reaction system).
Suppose we want to find the unknown total flux $F^*$ by extremizing the
corresponding total EP. However, this would be a meaningless operation, because
it would result in an infinity. We did not encounter this infinity in the
partial steady state MaxEP example discussed in section \ref{Variational MaxEP}, because
this principle did not use to the total EP, but a partial EP
instead. For the partial EP related with the non-linear processes, there is a
trade-off. One could have a non-fixed $X_k^*$ which might decrease when $F_k^*$
increases. Hence, due to this trade-off one avoids meaningless infinities.

Another point of discussion is the relation of the partial steady state MaxEP principle
with a transient (Lyapunov) principle. If the relaxation of the transient states towards the
steady state is such that the specific EP behaves as $\frac{d \sigma_{RW}}{d
t}\geq 0$, reaching its maximum in the selected steady state, then the
partial steady state MaxEP is a Lyapunov principle. It is an open question whether the
examples of the partial steady state MaxEP principle discussed in the literature have a
Lyapunov-type behavior. 

As we have presented a general counter example, this does not totally degrade
the value of MaxEP. The non-linear partial steady state MaxEP might be wrong in most
interesting cases (such as turbulent fluid flows \cite{CastilloHoover}), but on
the other hand, the partial steady state principle has some experimental strength
\cite{LorenzLunine} (see \cite{OzawaOhmura} for review). The latter principle is
not trivial, and its experimental verification specifically comes from
interesting systems, such as turbulent atmospheric and ocean systems. This
experimental corroboration needs some explanation. If one could find a
non-trivial theoretical model (an attempt was made in e.g. \cite{Paltridge2001})
that shows this partial steady state MaxEP, this might increase our understanding of some
highly non-linear systems.

\section*{Acknowledgments}

The author wishes to thank F. Meysman and C. Maes for helpful comments.


\end{document}